\documentclass[twocolumn,pra,amsmath,letterpaper,floatfix,showpacs,nolongbibliography]{revtex4-2}
\usepackage{graphicx}
\usepackage{amsmath}
\usepackage{amssymb}
\usepackage{color}
\usepackage{gensymb}
\usepackage{siunitx}

\newcommand{\otwo}[0]{$\mathrm{O_2}$}
\newcommand{\ntwo}[0]{$\mathrm{N_2}$}

\newcommand{\II}{{\mathbb{I}}}
\newcommand{\CtwoT}[0]{\langle \cos^2\theta_\text{2D} \rangle}

\begin{document}

\title{An optical centrifuge with zero rotational acceleration}

\author{K. Wang}
\author{A. Hardikar}
\author{A. A. Milner}
\author{V. Milner}
\email{vmilner@phas.ubc.ca}
\affiliation{Department of  Physics \& Astronomy, The University of British Columbia, Vancouver, Canada}

\date{\today}

\begin{abstract}
        An optical centrifuge is a laser pulse which enables controlled rotational excitation of molecules. Centrifuged molecules rotating with well-defined angular frequencies are ideal candidates to probe many-body quantum systems at the nanoscale. Because the interaction with the quantum environment increases the effective moment of inertia of the embedded molecular probes, the required rotational acceleration of the optical centrifuge must be lowered to accommodate adiabatic spinning. We demonstrate a new design of an optical centrifuge, based on the method of spectral focusing, which enables extremely low, down to zero, rotational accelerations. We discuss the potential of using such constant-frequency centrifuge to investigate molecular rotation inside helium nanodroplets.
    \end{abstract}

\maketitle

%----------- Introduction 1: Conventional optical centrifuge --------
Since its invention in 1999 \cite{Karczmarek1999, Villeneuve2000}, the optical centrifuge has become a powerful tool for controlling molecular rotation. The centrifuge is a linearly polarized laser pulse, whose polarization vector rotates with a constant angular acceleration. The interaction of the laser-induced dipole moment in the molecule of interest with the applied laser field results in a torque, which pushes the most polarizable molecular axis toward the field polarization, forcing the molecule to follow its accelerated rotation.

Optical centrifuges have been used in multiple experimental studies of molecular ro-vibrational structure \cite{Korobenko2014b, Milner2017b, Chen2023}, magnetic and optical properties of molecules undergoing unidirectional rotation \cite{Milner2014c, Faucher2016, Milner2021, Tutunnikov2022}, molecular collisions \cite{Yuan2011, Toro2013, Korobenko2014a, Milner2014a, Murray2018}, and the dynamics of molecular rotors in external fields \cite{Korobenko2015a, Milner2017a} (for recent reviews on this broad topic, see Refs.~\citenum{MacPhail2020, Mullin2025}). Aside from the study of centrifuged chiral molecules \cite{Milner2019}, the centrifuge has been applied to relatively small (diatomic and triatomic) molecular species in the gas phase. Because the efficiency of an optical centrifuge to rotationally excite any given molecule scales proportionally to the molecular polarizability anisotropy ($\Delta \alpha $) and inversely proportional to both the molecular moment of inertia ($\II$) and the angular acceleration of the centrifuge ($\epsilon  $), successful spinning of larger molecules with higher values of $\II/\Delta \alpha $ requires a slower centrifuge \cite{MacPhail2020}.

%----------- Introduction 2: Molecules in He nanodroplets -----------
High moments of inertia may stem not only from the increased complexity of the molecular structure, but also from the interaction between the molecular rotor and the surrounding environment. For example, from the studies of  molecules embedded in helium nanodroplets, both in the frequency \cite{Grebenev2000, Choi2006, Lehnig2009} and the time domains \cite{Pentlehner2013, Chatterley2020}, the molecular moment of inertia is known to increase by up to a factor of ~5 due to the weak bonding of He atoms to the solvated molecule. An even bigger effect has been observed in the centrifugal distortion constant of the ``molecule-He'' complex, which increases by a few orders of magnitude, setting an upper limit for the rotational states accessible by the optical centrifuge \cite{Cherepanov2021} (this effect, dubbed ``the centrifugal wall'', has also been studied in gas samples of molecular super-rotors \cite{Milner2017b}).

%----------- Introduction 3: Main goal of work ----------------------
The effects of helium on the solvated molecule (both to increase its moment of inertia and to decrease the accessible rotational frequencies) call for new optical tools to probe superfluidity at the nanoscale with molecular rotors. To utilize the proven controllability of rotational excitation by an optical centrifuge, it would therefore be imperative to decrease both its angular acceleration and its spectral bandwidth. However, reaching this goal with the original centrifuge design, based on a femtosecond pulse shaper \cite{Villeneuve2000}, becomes increasingly difficult. Both lowering the frequency chirp $\beta $ (responsible for the centrifuge acceleration $\varepsilon = 2\beta $) and shaping pulses with reduced spectral bandwidth $\Delta \omega $ (responsible for the rotational frequency $\Omega = \Delta \omega /2$) requires impractical shaper parameters (for details, see Ref.~\citenum{MacPhail2020}).

Here, we describe our recent development of an alternative method to generate the field of an optical centrifuge, which allows arbitrarily small values of $\left| \varepsilon \right|$ and $\left| \Omega \right|$. We start with the case of a ``constant-frequency centrifuge'' (cfCFG), for which $\varepsilon = 0$ and $\Omega /2\pi$ is as low as a few GHz. We outline the parameters of the new cfCFG  and demonstrate its performance in three different experimental scenarios: frequency-resolved rotational Raman spectroscopy, time-resolved detection of molecular alignment in a room-temperature molecular gas, and time-resolved detection of molecular alignment in a cold molecular jet.

\begin{figure*}[t]
    \includegraphics[width=2\columnwidth]{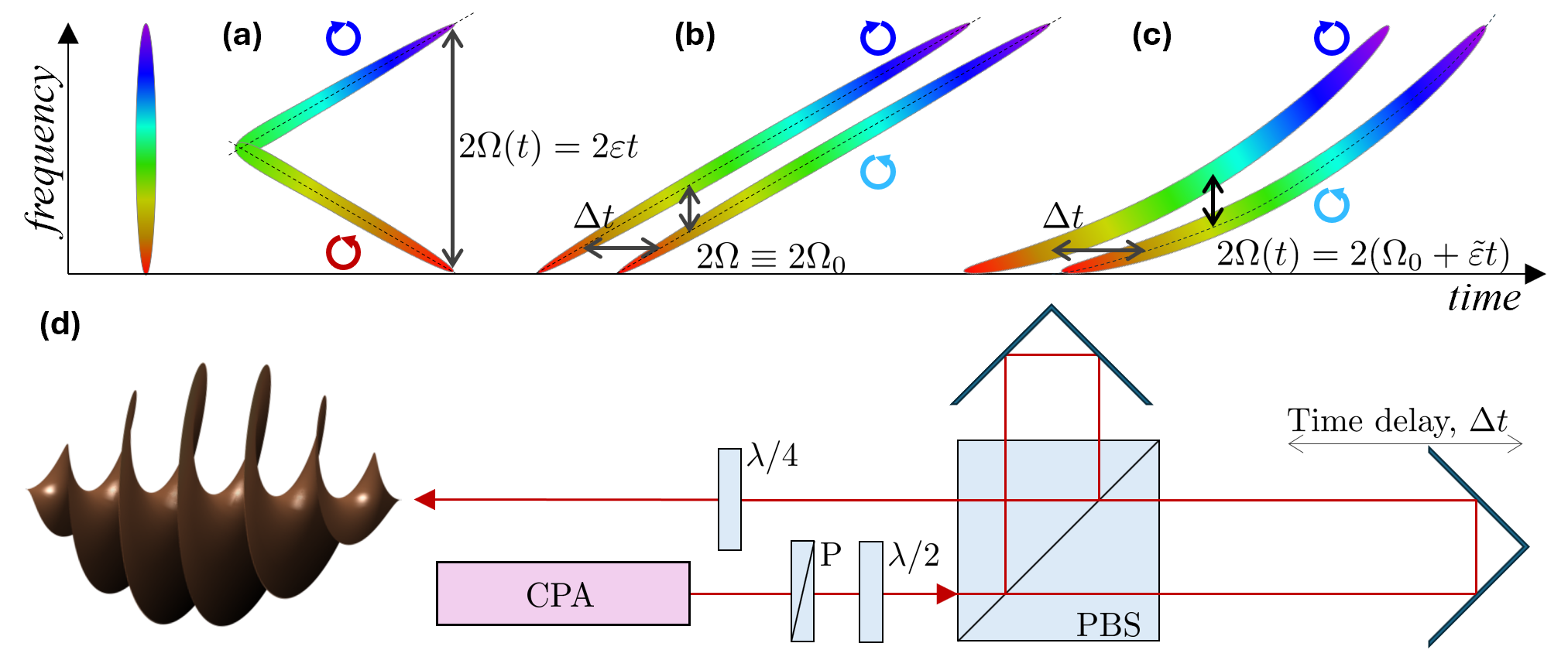}
    \caption{(color online) Time-frequency diagrams of laser pulses, corresponding to (\textbf{a}) conventional optical centrifuge and (\textbf{b}) constant-frequency optical centrifuge (cfCFG). Circles with arrows represent the polarization of the chirped pulses (tilted ellipses), whose interference results in the centrifuge field, rotating with frequency $\Omega(t) $. (\textbf{c}) Effects of third order dispersion (TOD, exaggerated for illustration purposes) on the time-frequency diagram of cfCFG. See text for the definition of $\Omega_0$, $\varepsilon$ and $\tilde{\varepsilon}$. (\textbf{d}) An outline of an optical setup producing the field of the cfCFG, illustrated with a corkscrew shape on the left. CPA: chirped pulse amplifier, P: linear polarizer, PBS: polarizing beamsplitter, $\lambda /2$ and $\lambda /4$: half- and quarter-wave plates.}
    \label{Fig-Chirps}
\end{figure*}

%----------- Setup 1: cf Centrifuge setup ---------------------------
The operational principle of a conventional centrifuge is illustrated in Fig.~\ref{Fig-Chirps}(\textbf{a}). A spectrally broad femtosecond laser pulse (schematically shown on the left side of Fig.~\ref{Fig-Chirps}) is split in half in the \textit{frequency domain}, and the two spectral components are frequency chirped with chirp rates of equal magnitude and opposite signs ($\pm \beta $). If the two fields are then circularly polarized with opposite handedness and overlapped in space and time, their interference produces a linearly polarized pulse, whose polarization vector rotates with an angular frequency $\Omega (t)=\varepsilon t$, linearly accelerating in time with acceleration $\varepsilon = 2\beta$ \cite{MacPhail2020}.

Here we propose an alternative design of an optical centrifuge, based on the concept of ``spectral focusing'', first introduced for high-resolution spectroscopy with broadband laser pulses \cite{Hellerer2004}. As illustrated in Fig.~\ref{Fig-Chirps}(\textbf{b}), if a frequency chirped pulse is split in half in the \textit{time domain} and the two pulses are delayed with respect to one another, the instantaneous frequency difference between them is constant and linearly proportional to the product of the frequency chirp and the time delay. As with the conventional centrifuge, overlapping the two fields in space and polarizing them with an opposite sense of circular polarization results in linear polarization, rotating at a constant angular frequency $\Omega_0 $. This is the pulse we refer to as the constant-frequency centrifuge.

%----------- Setup 2: Chirp and TOD ---------------------------------
Note that a small amount of third-order dispersion (TOD), typically present in broadband laser pulses, introduces angular acceleration to the rotating centrifuge polarization. Consider the following optical phase and the corresponding instantaneous frequency of the initial laser pulse:
\begin{equation}\label{phase}
    \varphi (t)=\varphi_0+\omega_0 t+\beta  t^2 +\gamma t^3,
\end{equation}
\begin{equation}\label{omega_inst}
    \omega (t)=\dot{\varphi}(t)=\omega_0+2\beta t +3\gamma t^2,
\end{equation}
\noindent where $\omega_0$ is the optical carrier frequency, $\beta $ is the frequency chirp rate, and $\gamma $ is the TOD parameter. As illustrated in Fig.~\ref{Fig-Chirps}(\textbf{c}), the frequency difference between the two interfering circularly polarized fields is no longer constant, but rather changes linearly with time, which results in the accelerated rotation of the polarization vector,
\begin{equation}\label{Omega_cfg}
    \Omega(t) = \frac{1}{2} \frac{\text{d} \omega(t)}{\text{d} t} \Delta t = \Omega_0 +\tilde{\varepsilon} t,
\end{equation}
\noindent with the average angular frequency and acceleration defined, respectively, as
\begin{equation}\label{TOD_acceleration}
    \Omega_0=\beta \Delta t,\text{ and } \tilde{\varepsilon} = 3\gamma \Delta t,
\end{equation}
\noindent where $\Delta t$ is the time delay between the arms of the interferometer. The effects of $\tilde{\varepsilon}$ on the performance of cfCFG will be discussed later in the text.

%----------- Setup 3: Michelson interferometer ----------------------
In contrast to a femtosecond pulse shaper in an optical setup of a conventional centrifuge, a cfCFG is produced by means of a Michelson interferometer with orthogonally polarized arms, as shown in Fig.~\ref{Fig-Chirps}(\textbf{d}). The output of a chirped pulse amplifier (CPA) is split with a polarizing beamsplitter (PBS) into two linearly polarized beams of equal intensity (balanced with a $\lambda /2$ plate). After the delay $\Delta t$ is added to one of the interferometer's arms, the two beams are combined and passed through a quarter-wave plate ($\lambda /4$), oriented at $45^\circ$ with respect to the PBS.

\begin{figure}[t]
    \includegraphics[width=1\columnwidth]{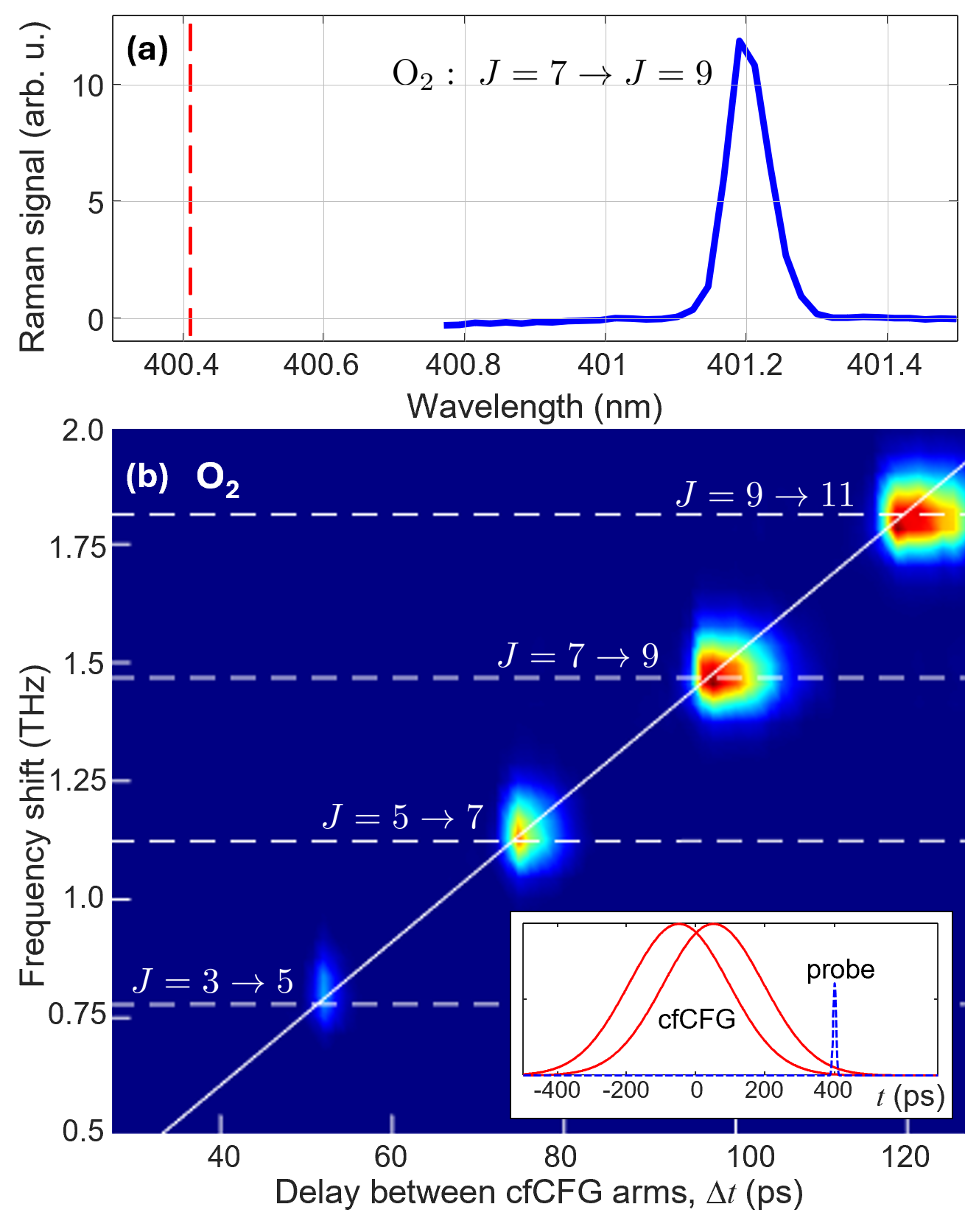}
    \caption{(color online) Coherent Raman scattering from the gas of oxygen molecules, rotationally excited by the constant-frequency centrifuge. (\textbf{a}) Example of a Raman signal obtained with the cfCFG tuned to the transition between $J=7$ and $J=9$ rotational levels of O$_{2}$. Red vertical dashed line indicates the central wavelength of probe pulses. (\textbf{b}) Raman spectrogram, recorded by scanning the time delay $\Delta t$ between the two arms of the centrifuge. Horizontal dashed lines, labeled with the rotational Raman resonances of oxygen molecule, indicate the calculated frequencies of the corresponding Raman shifts. Diagonal solid line illustrates the predominantly linear dependence of the Raman shift with respect to $\Delta t$. Inset in the lower right corner outlines the relative timing between the two centrifuge arms (solid red) and a probe pulse (dashed blue).}
    \label{Fig-Raman}
\end{figure}
%----------- Experiment 1: Calibration with Raman -------------------
After implementing the design of a constant-frequency centrifuge described above, the calibration of its frequency as a function of the time delay $\Delta t$ was carried out using two complementary methods. First, we employed the technique of coherent Raman scattering. Owing to the centrifuge-induced coherence between the quantum states separated by $\Delta J=\pm2$ (where $J$ is the rotational quantum number), the spectrum of a probe light passing through the centrifuged ensemble develops Raman sidebands. The magnitude of the Raman shift equals twice the rotation frequency, while its sign reflects the direction of molecular rotation with respect to the probe circular polarization \cite{Korobenko2014a}.

The experiments on Raman scattering were done in oxygen gas at room temperature and a pressure of \SI{0.5}{bar}. The experimental setup and procedures have been discussed in detail in our previous publications on rotational excitation of molecules with a conventional optical centrifuge \cite{Korobenko2014b, Milner2014a, Milner2014b}. Briefly, feeding the interferometer with the uncompressed output of Spectra-Physics Spitfire Ace (total pulse energy of 2~mJ) produced centrifuge pulses, which were then focused in the gas cell by a lens with a focal length of 10~cm. Probe pulses were extracted from the same laser system (compressed output, total pulse energy of \SI{2.5}{mJ}), spectrally narrowed to 0.1~nm in a $4f$ pulse shaper, and frequency doubled in a BBO crystal to enable convenient separation from the centrifuge light. They were then delayed with respect to the cfCFG pulses by 400~ps (see inset in the lower right corner of  Fig.~\ref{Fig-Raman}), and sent to a single-grating spectrometer (McPherson 2035).

An example of the probe spectrum, recorded with the delay between cfCFG arms of $\Delta t=\SI{100}{ps}$, is shown in Fig.~\ref{Fig-Raman}(\textbf{a}). Here, the observed Raman peak is shifted by about \SI{0.8}{nm} (or \SI{1.5}{THz}) from the central probe wavelength (vertical red line). It corresponds to the $J=7\rightarrow 9$ Raman transition in molecular oxygen, driven by the centrifuge field. As the time delay between the two chirped pulses of the cfCFG (i.e. the arms of the Michelson interferometer, see Fig.~\ref{Fig-Chirps}) was scanned between 30 and 130~ps, we detected a series of such Raman resonances, shown in Fig.~\ref{Fig-Raman}(\textbf{b}), which are in good agreement with the calculated frequencies (horizontal dashed lines, \cite{NIST}). From the linear fit of the observed Raman peaks (diagonal solid line), one can extract the calibration parameter of the centrifuge $\Omega /\Delta t = 2\pi\times\SI{7.5\pm0.5}{rad\,GHz/ps}$.

\begin{figure}[t]
    \includegraphics[width=1\columnwidth]{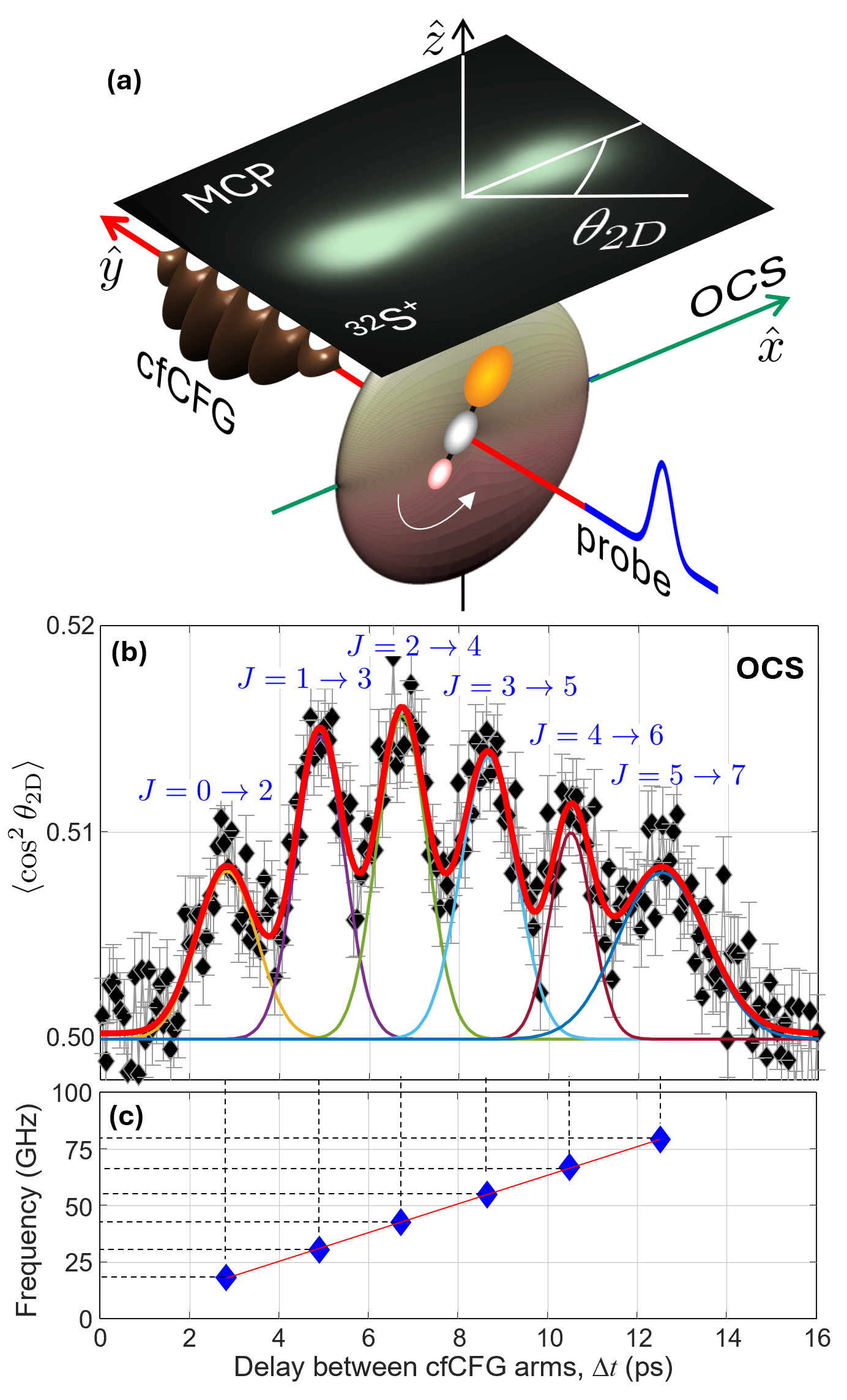}
    \caption{(color online) (\textbf{a}) Geometry of the experiment with OCS molecules in seeded molecular jet using the technique of velocity map imaging with a multi-channel plate (MCP) detector. (\textbf{b}) Degree of planar alignment of OCS molecules, rotationally excited by the constant-frequency centrifuge, as a function of the time delay $\Delta t$ between the centrifuge arms. Solid red line is the best fit of experimental data (grey dots with error bars) to the sum of six gaussian peaks, plotted with colored curves and labeled with the corresponding excitation transition. (\textbf{c}) Rotational frequency of cfCFG, $\Omega/2\pi $, extracted from the known transition frequencies of OCS (blue diamonds). Solid red line shows the best fit to the linear dependence on $\Delta t$.}
    \label{Fig-VMIcalib}
\end{figure}
%----------- Experiment 2: Calibration with VMI ---------------------
To carry out the same calibration at a much lower frequency range ($\ll \SI{1}{THz}$, where a constant-frequency optical centrifuge is expected to be most useful), we switched to a heavier molecule, carbonyl sulfide (OCS). To increase the population of lower rotational states, the experiments were conducted in a molecular jet of OCS seeded in helium at a concentration of \SI{5000}{ppm}. Expanding the gas mixture into a vacuum chamber through a pulsed nozzle with a backing pressure of \SI{7}{bar} results in the rotational temperature of OCS of \SI{\sim 3.5}{K}. At this temperature, most molecules occupy one of the six lowest rotational states, $J=0...5$, corresponding to the rotational frequencies between about \SI{20}{GHz} and \SI{80}{GHz}. The molecules were excited by a cfCFG, whose constant frequency was scanned in this low-frequency window by varying the delay between the centrifuge arms.

We employed velocity map imaging (VMI) \cite{Eppink1997}, a technique often used in studies of molecular rotation \cite{Stapelfeldt2003}, to detect the degree of rotational excitation. The centrifuge pulse (\SI{\approx 330}{ps} FWHM, \SI{3.9}{mJ}, \SI{\approx 2e12}{W/cm^2}) is followed by an intense femtosecond probe pulse (\SI{\approx 120}{fs}, \SI{150}{\mu J}, \SI{\approx 2e14}{W/cm^2}), which causes multiple ionization of OCS. This triggers an immediate and near-instantaneous Coulomb explosion of the molecules into positively charged ion fragments. A conventional time-of-flight (TOF) setup accelerates the fragments toward a multi-channel plate (MCP), gated for the arrival of a fragment of interest with a particular mass-to-charge ratio (here, $^{32}$S$^{+}$). Each ion hit is recorded by means of a phosphor screen and a camera, and its location in the MCP plane is quantified by the angle $\theta_\text{2D}$ with respect to the rotational plane of the centrifuge, as illustrated in Fig.~\ref{Fig-VMIcalib}(\textbf{a}). Probe pulses are linearly polarized along $\hat{z}$, which eliminates any anisotropy of ion images in the $xy$ plane due to the ionization process.

Averaged over the molecular ensemble, the deviation of $\CtwoT{}$ from its isotropic value of 0.5 is a convenient measure of the degree of rotational excitation \cite{Soendergaard2017}. The observed dependence of $\CtwoT{}$ on the delay between the centrifuge arms $\Delta t$, and hence on the centrifuge frequency, is shown in Fig.~\ref{Fig-VMIcalib}(\textbf{b}). We identified the detected peaks in $\CtwoT{}$ as the first six rotational resonances of OCS, driven by cfCFG via two-photon Raman transitions, indicated by the corresponding labels in the figure.

Taking the known values for the transition frequencies \cite{NIST}, we plot them as a function of $\Delta t$ in Fig.~\ref{Fig-VMIcalib}(\textbf{c}). An apparent linear dependence (solid red line) provides us with the calibration parameter of the centrifuge $\Omega /\Delta t = 2\pi\times\SI{6.4\pm0.2}{rad\,GHz/ps}$. The difference between this calibration factor and that found by means of coherent Raman scattering in centrifuged oxygen is related to third-order dispersion. Indeed, from Eqs.~(\ref{Omega_cfg}, \ref{TOD_acceleration}), it follows that
\begin{equation}\label{Calibration}
    \Omega (t)/\Delta t=\beta +3\gamma t.
\end{equation}
\noindent For a given resonant frequency $\Omega_\text{res}$ and a fixed cfCFG-probe delay, the strongest signal is expected when the centrifuge reaches $\Omega_\text{res}$ at its peak intensity, i.e. when $\Omega (t=0) = \Omega_\text{res}$ (see inset in Fig.~\ref{Fig-Raman}). This is indeed true for $\CtwoT{}$, which does not decay on the timescale of the experiment, owing to the lack of collisions in the molecular jet. However, due to the collision-induced decay of the Raman signal \cite{Milner2014a}, it will reach maximum when the centrifuge becomes resonant with the Raman transition at a later time, $t>0$, closer to the probe pulse. According to Eq.~(\ref{Calibration}), this will increase the $\Omega (t)/\Delta t$ ratio, in agreement with our observations. Similarly, TOD is also responsible for the finite width of the resonant VMI peaks, around $\SI{1.5}{ps}$ in Fig.~\ref{Fig-VMIcalib}(\textbf{b}), which defines the resolution limit of $\approx\SI{10}{GHz}$ in this frequency range.

%------------ Results 1: N2 rotation with VMI -----------------------
In the above analysis, we have studied \textit{field-free} molecular rotation, initiated by the constant-frequency centrifuge. Much like the conventional centrifuge, cfCFG drives resonant two-photon Raman transitions between the molecular $J$ states, creating rotational wave packets which persist after the centrifuge pulse. Naturally, the only allowed frequencies of such field-free rotation are determined by the rotational spectrum of the molecule, which we used to calibrate the centrifuge.

On the other hand, the much more compact design of cfCFG, based on the interferometrically stable Michelson setup, allowed us to observe \textit{in-field} molecular rotation with arbitrary frequencies. Indeed, a constant phase difference between the two arms of the interferometer (typically unattainable with a conventional centrifuge) means that the orientation of the corkscrew centrifuge field (see Fig.~\ref{Fig-Chirps}), i.e. the angle of its polarization vector at any given time, does not change from one laser pulse to another. As a result, not only the frequency but also the \textit{phase} of the centrifuge-induced molecular rotation is shot-to-shot stable, enabling one to follow this rotation in time without the random phase averaging. We demonstrate this new feature of an optical centrifuge with two examples below.

\begin{figure}[t]
    \includegraphics[width=1\columnwidth]{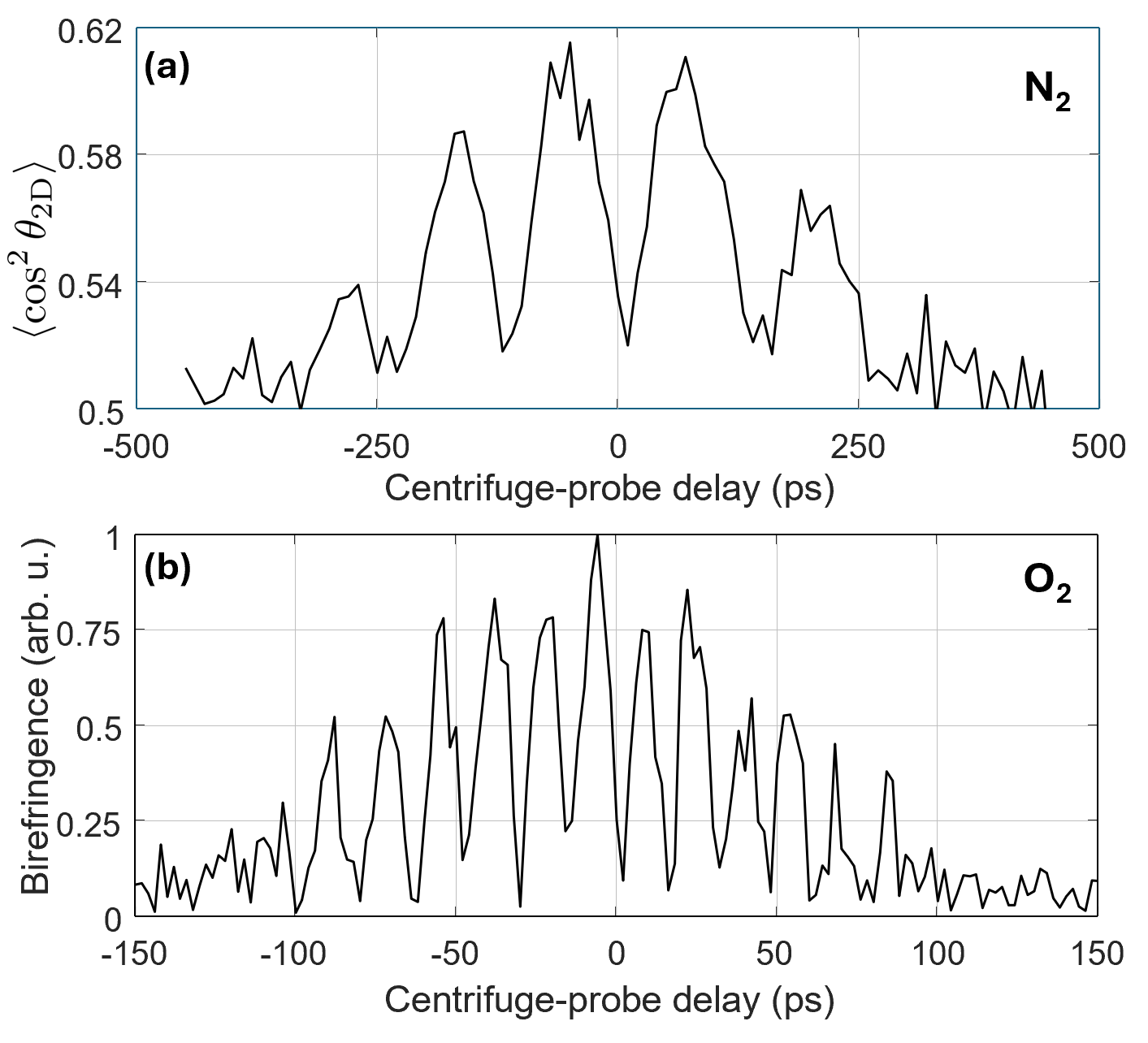}
    \caption{(\textbf{a}) Ensemble-averaged alignment factor $\CtwoT{}$ of nitrogen molecules in a molecular jet, exposed to the field of a constant-frequency centrifuge. See text and Fig.~\ref{Fig-VMIcalib} for details on the VMI-based detection technique. (\textbf{b}) Centrifuge-induced optical birefringence of a dense gas of oxygen molecules at room temperature. See text for the details of the polarimetry-based detection method.}
    \label{Fig-insideCFG}
\end{figure}

In the first example, nitrogen molecules were expanded in the vacuum chamber through a pulsed nozzle, irradiated by cfCFG pulses, then Coulomb exploded similarly to the experiments with OCS, described above. However, instead of sending the probe at a constant delay after the centrifuge pulse and scanning the cfCFG frequency, we scanned the probe delay across the centrifuge, whose rotation frequency $\Omega $ has been kept fixed. The MCP detector was gated for the arrival of N$^+$ ions. The observed $\CtwoT{}$ is plotted in Fig.~\ref{Fig-insideCFG}(\textbf{a}) for $\Omega \approx2\pi\times\SI{4}{rad\,GHz}$.

Clear oscillations at $\SI{8}{GHz}$, corresponding to the angular frequency of $2\Omega $, confirm that \ntwo{} molecules are following the rotation of the linearly polarized centrifuge field. Indeed, as the molecules rotate in the $xz$ plane, the value of $\CtwoT{}$ alternates between 0.5, when molecular axes are perpendicular to the plane of the MCP detector [$\hat{z}$ in Fig.~\ref{Fig-VMIcalib}(\textbf{a})], and a maximum value at the time of molecular alignment along $\hat{x}$ (both occurring twice per rotation period). Similar to the well-known adiabatic molecular alignment by a non-rotating intense linearly polarized laser pulse \cite{Friedrich1995}, the rotating alignment observed here stems from the field-induced pendular states. Incidentally, the highest value of $\CtwoT{}$ (about 0.6 here) is determined by multiple factors, such as the strength of the laser field, the molecular polarizability anisotropy, and the temperature of the molecular ensemble \cite{Stapelfeldt2003}.

%------------ Results 2: O2 rotation with birefringence -------------
In the second example, we show that in-field forced molecular rotation can also be established in dense gas samples. Here, we applied cfCFG of shorter duration (\SI{120}{ps} FWHM) to the room-temperature oxygen gas under a pressure of \SI{0.5}{bar}. To follow the rotating axes of \otwo{} molecules, we employed the polarimetry-based technique to detect the alignment-induced optical birefringence \cite{Renard2004}, outlined below.

Centrifuge pulses (\SI{4e12}{W/cm^2}) were focused in the center of a gas cell, and were followed by linearly polarized ultrashort probe pulses (\SI{70}{fs} FWHM) at a varied time delay. A linear polarizer, oriented at 90 degrees with respect to the input probe polarization, was installed in front of the photodiode detector. The rotating alignment of oxygen molecules, following the rotation of the centrifuge field, leads to the time-dependent optical birefringence of the gas. This results in the periodic change of probe polarization from linear (when molecular axes are lined up either along with, or perpendicular to, the input probe polarization) to slightly elliptical at any other alignment angle (with the maximum ellipticity at the angles of $\pm45^{\circ}$ and $\pm135^{\circ}$).

The experimentally observed periodic oscillations of the centrifuge-induced birefringence of \otwo{} are shown in Fig.~\ref{Fig-insideCFG}(\textbf{b}). Using the calibration of the centrifuge, described earlier in the text, the delay between the centrifuge arms was set to produce the rotation at the frequency of $\Omega =2\pi\times\SI{15.5\pm0.5}{GHz}$. As the birefringence peaks are reached four times during a single rotation period, the observed frequency of $4\times\SI{15.8\pm0.1}{GHz}$ is in good agreement with the expected value.

%-------------------Conclusion --------------------------------------
To summarize, we have introduced a new technique to produce a laser field with rotating linear polarization, known as an optical centrifuge. Our method expands the capabilities of the conventional centrifuge in two important aspects. First, it allows a significant reduction in the terminal frequency and the angular acceleration of the centrifuge down to much lower values. To that end, we have built and demonstrated the operation of a constant-frequency centrifuge, which rotates with nearly zero acceleration in the broad frequency range from a few gigahertz to a few terahertz. The acceleration can be reduced as low as required, down to zero, by removing the small amount of third-order dispersion \cite{Stern1992}, whose effects on the performance of cfCFG have been discussed.

Secondly, we have established that the phase stability of the new optical centrifuge results in the equally stable rotational phase of the centrifuged molecules. This means that the orientation of molecular axes at a given time within the envelope of the  centrifuge pulse does not change from one laser shot to another. The two unique properties of the constant-frequency optical centrifuge, summarized above, have been demonstrated in multiple molecular systems, from dense room-temperature oxygen to cold carbonyl sulfide in a seeded molecular jet, using three different detection techniques: coherent Raman spectroscopy, ultrafast polarimetry, and velocity map imaging. The utility of cfCFG for the controlled spinning of molecules embedded in the strongly-interacting quantum environment of helium nanodroplets has been recently demonstrated \cite{MacPhail2025}.

%---------------- Outlook: Adding acceleration with a chirp ---------
We are currently extending the reported centrifuge design to include low angular accelerations by adding a grating-based pulse stretcher to one of the centrifuge arms. The latter will introduce a small difference in the frequency chirp rates of the two interfering circularly polarized fields, allowing us to control the rotation of the linear polarization in an arbitrarily low range of accelerations and terminal frequencies. The details and performance of such an ultra-slow centrifuge \textit{with variable angular acceleration} will be discussed in the upcoming publication. We anticipate that the new type of optical centrifuge discussed in this work will provide a useful toolbox to probe many-body quantum systems with molecular rotors undergoing controlled unidirectional rotation.

%--------------Acknowledgement ----------------------------------------
We thank I.~MacPhail-Bartley for valuable comments on the manuscript. This research has been supported by the grants from CFI, BCKDF and NSERC and carried out under the auspices of the UBC Center for Research on Ultra-Cold Systems (CRUCS).

%\bibliography{cfCFG}


\end{document}